\documentclass{aa}  

\usepackage{graphicx}
\usepackage{txfonts}
\usepackage{multirow}
\usepackage[]{hyperref}
\begin{document} 

    \titlerunning{The aspherical explosions of the 03fg-like Type~Ia SNe 2021zny and 2022ilv revealed by polarimetry}  
    \authorrunning{T. Nagao et al.}

   \title{The aspherical explosions of the 03fg-like Type~Ia supernovae 2021zny and 2022ilv revealed by polarimetry}


   \author{T. Nagao, \inst{1,2,3} 
          K. Maeda, \inst{4} 
          S. Mattila, \inst{1,5} 
          H. Kuncarayakti, \inst{1,6} 
          C.~P.~Guti{\'e}rrez, \inst{7,8} 
          \and
          A. Cikota \inst{9}
          }

    \institute{
            Department of Physics and Astronomy, University of Turku, FI-20014 Turku, Finland
            \and
            Aalto University Mets\"ahovi Radio Observatory, Mets\"ahovintie 114, 02540 Kylm\"al\"a, Finland
            \and
            Aalto University Department of Electronics and Nanoengineering, P.O. BOX 15500, FI-00076 AALTO, Finland
            \and
            Department of Astronomy, Kyoto University, Kitashirakawa-Oiwake-cho, Sakyo-ku, Kyoto 606-8502, Japan
            \and
            School of Sciences, European University Cyprus, Diogenes Street, Engomi, 1516, Nicosia, Cyprus
            \and
            Finnish Centre for Astronomy with ESO (FINCA), University of Turku, FI-20014, Finland
            \and
            Institut d'Estudis Espacials de Catalunya (IEEC), Edifici RDIT, Campus UPC, 08860 Castelldefels (Barcelona), Spain
            \and
            Institute of Space Sciences (ICE, CSIC), Campus UAB, Carrer de Can Magrans, s/n, E-08193 Barcelona, Spain
            \and
            Gemini Observatory / NSF's NOIRLab, Casilla 603, La Serena, Chile\\
             }

   \date{Received ***; accepted ***}

 
  \abstract
   {A peculiar subtype of Type~Ia supernovae (SNe), 03fg-like (super-Chandrasekhar) SNe, show different observational properties from prototypical Type~Ia SNe, typically having high luminosity at the light-curve peak, low expansion velocities, and strong carbon features. The origin of this class of Type~Ia SNe has been actively debated. Recent nebular-phase infrared observations of the 03fg-like Type Ia SN~2022pul using the James Webb Space Telescope revealed large-scale asymmetries in the ejecta and the presence of the strong [Ne II] line at 12.81 $\mu$m, suggesting a violent merger of two white dwarfs as its origin.}
   {Polarimetry is another powerful tool to study overall ejecta asymmetries of spatially-unresolved SNe. Here, we aim to check the universality of the violent merger scenario as the origin of the 03fg-like Type Ia SNe, by studying their explosion geometries using polarimetry.}
   {In this letter, we present imaging-polarimetric observations of the two 03fg-like Type~Ia SNe 2021zny and 2022ilv.}
   {SNe~2021zny and 2022ilv show high intrinsic polarization ($\sim1$ \% -$\sim2$\%), which might be composed of multiple components with different polarization angles. This indicates that they have complex aspherical structures in their ejecta, supporting the violent merger scenario for their origin. Our observations provide the first clear evidence from polarimetry for such aspherical structures.
   }
   {}


   \maketitle

%

\section{Introduction}

Type~Ia supernovae (SNe) are explosions of white dwarfs (WDs), powered by runaway thermonuclear burning of the degenerate gas \citep[e.g.,][]{Maeda2016,Livio2018}. Given that the majority of Type~Ia SNe show standardizable light-curve properties, they are popular standard candles for cosmological distance measurements \citep[e.g.,][]{Riess1998,Perlmutter1999}. In addition to such standardizable ``normal" Type~Ia SNe, it has been recognized that there are sub-categories in Type~Ia SNe, which show a wide range of photometric and spectroscopic properties \citep[e.g.,][]{Taubenberger2017,Jha2019}. These observational diversities are coupled with the different progenitor systems, different evolutionary paths, and/or different thermonuclear burning behavior. However, we have not fully understood the progenitors and explosion mechanisms for different sub-types of Type~Ia SNe \citep[e.g.,][]{Maoz2014,Liu2023}.

One of such extremes in Type~Ia SNe is so-called 03fg-like (or super-Chandrasekhar) Type~Ia SNe, showing different observational properties from ``normal" Type~Ia SNe \citep[e.g.,][]{Howell2006,Hicken2007,Scalzo2010,Yamanaka2009,Taubenberger2019,Hsiao2020,Ashall2021,Dimitriadis2023,Srivastav2023}. They are typically brighter with a peak absolute $B$-band magnitude of $-19<M_{B}<-21$ mag \citep[][]{Ashall2021}. In some extreme cases, the required amount of $^{56}$Ni would be similar to the Chandrasekhar-mass ($\sim1.4$ M$_{\odot}$) and thus require super-Chandrasekhar-mass WDs as their progenitors, unless other energy sources significantly contribute to their brightness alike in normal Type~Ia SNe \citep[e.g.,][]{Yamanaka2009}. They show slow evolution in their light curves (LCs) with $\Delta m_{15}(B)<1.3$ mag as well as relatively low expansion velocities (8000–12000 km s$^{-1}$ for Si II $\lambda 6355$) and strong features from unburnt carbon in their spectra \citep[][]{Ashall2021}, suggesting more massive SN ejecta compared to normal Type~Ia SNe \citep[e.g.,][]{Taubenberger2017}.

There are several scenarios proposed for explaining the observational properties of the 03fg-like Type Ia SNe.
(1) An explosion of a carbon-oxygen (CO) WD whose mass exceeds the Chandrasekhar limit due to its rapid rotation \citep[e.g.,][]{Yoon2005} and/or strong magnetic fields \citep[e.g.,][]{Das2013}, formed through the accretion from a non-degenerate companion in the single-degenerate scenario \citep[e.g.,][]{Whelan1973,Nomoto1982} or through post-merger accretion in the double-degenerate scenario \citep[e.g.,][]{Iben1984,Webbink1984,Raskin2014}.
(2) An explosion of a CO WD during a merger with a companion WD \citep[the violent merger scenario; e.g.,][for a review]{Pakmor2017}. In this scenario, the high peak luminosity might be achieved due to viewing angle effects of an aspherical explosion \citep[e.g.,][]{Hillebrandt2007,Moll2014} and/or due to interaction with circumstellar material \citep[CSM; e.g.,][]{Hachinger2012,Noebauer2016}.
(3) An explosion after a merger of a WD with the core of a massive asymptotic giant branch star \citep[the core-degenerate scenario; e.g.,][]{Sparks1974,Livio2003,Kashi2011}. This scenario may also explain the large brightness with a CSM interaction.
It is noted that recent very early-phase (within a few days after the explosion) observations of several 03fg-like SNe have detected early excesses in their LCs, suggesting interaction with H-poor CSM \citep[SNe 2020hvf, 2021zny, 2022ilv and 2022pul;][]{Jiang2021,Dimitriadis2023,Srivastav2023,Siebert2024}. However, the strength of CSM interaction estimated from the observed early excesses is not sufficient to boost the peak brightness from the level of normal Type Ia SNe ($\sim -19$ mag) to the observed level in bright 03fg-like SNe \citep[$\sim -20$ mag; e.g.,][]{Jiang2021,Maeda2023}. Therefore, bright 03fg-like SNe still need an additional stronger CSM interaction or super-Chandrasekhar mass $^{56}$Ni as the origin of their extreme brightness.

Nebular-phase observations of the 03fg-like Type Ia SN~2022pul with the James Webb Space Telescope exhibited anti-correlated asymmetric emission-line profiles for the iron-group elements (Fe, Co, Ni) and the intermediate-mass elements (S, Ar, Ca), as well as the presence of strong [Ne II] at 12.81 $\mu$m \citep[][]{Kwok2023,Siebert2024}. The separate distributions of different elements suggest large global ejecta asymmetries in SN~2022pul, and support the violent merger scenario. At the same time, the presence of the strong [Ne II] line was also predicted as a proof of a violent merger scenario by non-local-thermodynamic-equilibrium radiative transfer calculations for various scenarios in Type Ia SNe \citep[][]{Blondin2023}. These discoveries on SN~2022pul strongly support the violent merger scenario for its origin.

Polarimetry provides another powerful way to study the ejecta geometries of SNe. Its application to the 03fg-like Type~Ia SNe has been very limited due to the rareness of this class of SNe, but can be the key to understanding their origin. 
Spectropolarimetric observations of the 03fg-like Type Ia SN 2009dc show low continuum polarization ($<0.3$ \%) and indicate that the explosion is nearly symmetric \citep[][]{Tanaka2010}. This suggests that the explosion mechanism of SN~2009dc is different from that of SN~2022pul, which showed largely aspherical ejecta. Alternatively, this might merely imply that SN~2009dc has a similarly aspherical explosion but with a different viewing angle, i.e., SN~2009dc might be viewed from a direction close to the axis of symmetry. In fact, an aspherical system is expected for SN~2009dc from the line shapes in the nebular spectra \citep[][]{Siebert2023}.
Another example is SN~2007if, which shows relatively high wavelength-independent polarization ($P\sim 0.7$ \%, $\theta\sim 130$ degrees) from 13 to 46 days after the brightness peak \citep[][]{Cikota2019,Chu2022}. \citet[][]{Chu2022} conclude that this polarization likely originates from the interstellar polarization in the Milky Way (MW) as suggested by the relatively high MW extinction, implying that the intrinsic SN polarization is low and thus the explosion should be relatively spherical or viewed from the axis of symmetry.
Here, it is noted that the polarization angle of the observed polarization in SN~2007if is not similar with the directions of the interstellar polarization (ISP) of MW stars and the interstellar magnetic fields towards nearby directions from the SN line of sight \citep[$\sim 40$ degrees,][]{Berdyugin2014,Bennett2013}, although the local values do not necessarily follow the global values.

In this paper, we present imaging-polarimetric observations of two 03fg-like Type~Ia SNe: SNe 2021zny and 2022ilv \citep[][]{Yamanaka2021,Burke2022}.
SN~2021zny showed several characteristics of the class of 03fg-like Type~Ia SNe, such as high peak brightness, slow LC evolution, low ejecta velocities and strong lines from unburnt material \citep[][]{Dimitriadis2023}. \citet[][]{Dimitriadis2023} also report the detection of a flux excess within a few days after the explosion, which can be explained by interaction of the ejecta with $\sim 0.04$ M$_{\odot}$ of circumstellar material at a distance of $\sim 10^{12}$ cm, and prominent [O I] $\lambda\lambda$ 6300, 6364 lines at a late phase. From these observational properties, \citet[][]{Dimitriadis2023} conclude that the origin of SN~2021zny is possibly a merger of two CO WDs, where the disrupted secondary WD ejects carbon-rich CSM before the explosion of the primary WD.
\citet[][]{Srivastav2023} demonstrated that SN~2022ilv showed similar photometric and spectroscopic features as 03fg-like Type~Ia SNe as well as an early excess in the LC, and also proposed a similar merger scenario as its origin.

%
%
%

\section{Observations}

We conducted $V$- and $R$-band imaging polarimetry using the Alhambra Faint Object Spectrograph and Camera (ALFOSC)\footnote{\url{http://www.not.iac.es/instruments/alfosc/}} on the 2.56m Nordic Optical Telescope (NOT\footnote{\url{http://www.not.iac.es}}) for 03fg-like Type~Ia SNe~2021zny and 2022ilv. The observing logs are shown in Tables \ref{tab:log_21zny} and \ref{tab:log_22ilv}. For the linear polarimetry of the SNe, we utilized a half-wave plate (HWP) and a calcite plate. 
The HWP rotates the polarization axis of the transient light with a certain amount of angle, and then the transient light is split by a calcite plate into two orthogonally polarized beams (the ordinary and extraordinary components). We derived the Stokes parameters from the signals of the ordinary and extraordinary components for 4 HWP angles ($0^{\circ}$, $22.5^{\circ}$, $45^{\circ}$ and $67.5^{\circ}$).

The data were reduced and analyzed by the standard methods, e.g., in \citet[][]{Patat2017}, using IRAF\citep[][]{Tody1986,Tody1993}. First, we applied the basic treatment (cosmic-ray removal, bias and flat-field corrections) to all the frames. Then, we performed aperture photometry on the ordinary and the extraordinary components of the transient for all the HWP angles. 
Since the ordinary and extraordinary beams are overlapped in the ALFOSC images, an artificial polarization signal due to the inhomogeneous structures of the host galaxy and/or the background region can be created. In order to assess such an error for the polarization, we conducted aperture photometry using four different combinations of the aperture size and sky region: 
(1) an aperture size twice as large as the full width at half maximum (FWHM) of the ordinary beam's point-spread function with a sky region whose inner and outer radii are twice and three times as large as the FWHM; (2) an aperture size twice as large as the FWHM with a sky region whose inner and outer radii are three times and four times as large as the FWHM; (3) an aperture size 2.5 times as large as the FWHM with a sky region whose inner and outer radii are 2.5 times and 3.5 times as large as the FWHM; (4) an aperture size 2.5 times as large as the FWHM with a sky region whose inner and outer radii are 3.5 times and 4.5 times as large as the FWHM. Based on the measurements from the aperture photometry of the ordinary and extraordinary sources for 4 different HWP angles, we derived the Stokes $q$ and $u$ values for each combination of the aperture and sky region: ($q_1 \pm \sigma_{q,1}$,$u_1 \pm \sigma_{u,1}$), ($q_2 \pm \sigma_{q,2}$,$u_2 \pm \sigma_{u,2}$), ($q_3 \pm \sigma_{q,3}$,$u_3 \pm \sigma_{u,3}$), ($q_4 \pm \sigma_{q,4}$,$u_4 \pm \sigma_{u,4}$). Here, $\sigma$ represents the photon shot noise. Then, we took the average and standard deviation of these $q$ and $u$ values as the polarization signal and the error, respectively:
\begin{eqnarray}
 q_{\rm{ave}} = \frac{\Sigma^{n}_{i=1} \left( \frac{q_i}{\sigma_{q,i}^2} \right)}{\Sigma^{n}_{i=1} \left( \frac{1}{\sigma_{q,i}^2} \right)}, \sigma_{q,\rm{ave}}= \sqrt{\frac{\Sigma^{n}_{i=1} \left( \frac{1}{\sigma_{q,i}^2} \right)(q_i - q_{\rm{ave}})^2}{(n-1) \Sigma^{n}_{i=1} \left( \frac{1}{\sigma_{q,i}^2} \right)}},\\
 u_{\rm{ave}} = \frac{\Sigma^{n}_{i=1} \left( \frac{u_i}{\sigma_{u,i}^2} \right)}{\Sigma^{n}_{i=1} \left( \frac{1}{\sigma_{u,i}^2} \right)}, \sigma_{u,\rm{ave}}= \sqrt{\frac{\Sigma^{n}_{i=1} \left( \frac{1}{\sigma_{u,i}^2} \right)(u_i - u_{\rm{ave}})^2}{(n-1) \Sigma^{n}_{i=1} \left( \frac{1}{\sigma_{u,i}^2} \right)}}.
\end{eqnarray}
Here, $n$ is the number of the measurements to be avaraged and $n=4$.
From these averaged $q$ and $u$ values, we calculated the polarization degree and the polarization angle:
\begin{eqnarray}
P &=& \sqrt{q_{\rm{ave}}^{2} + u_{\rm{ave}}^{2}},\\ 
\sigma_{P} &=& \sqrt{\left( \frac{\partial P}{\partial q_{\rm{ave}}} \sigma_{q,\rm{ave}} \right)^2 + \left( \frac{\partial P}{\partial u_{\rm{ave}}} \sigma_{u,\rm{ave}} \right)^2} \nonumber\\
&=& \sqrt{\left( \frac{q_{\rm{ave}}}{P} \sigma_{q,\rm{ave}} \right)^2 + \left( \frac{u_{\rm{ave}}}{P} \sigma_{u,\rm{ave}} \right)^2},\\
\chi &=& \frac{1}{2} \arctan \left( \frac{u_{\rm{ave}}}{q_{\rm{ave}}} \right),\\
\sigma_\chi &=& \sqrt{\left( \frac{\partial \chi}{\partial q_{\rm{ave}}} \sigma_{q,\rm{ave}} \right)^2 + \left( \frac{\partial \chi}{\partial u_{\rm{ave}}} \sigma_{u,\rm{ave}} \right)^2} \nonumber\\
&=& \frac{\sqrt{ \left( u_{\rm{ave}} \sigma_{q,\rm{ave}} \right)^2 + \left( q_{\rm{ave}} \sigma_{u,\rm{ave}} \right)^2}}{2P^2}.
\end{eqnarray}
At last, we subtracted the polarization bias from the polarization degrees following the standard method in \citet[][]{Wang1997}.

\begin{table*}
      \caption[]{Log of the imaging-polarimetric observations of SN 2021zny and their measurements.}
      \label{tab:log_21zny}
      $
         \begin{array}{lcccccccccc}
            \hline
            \noalign{\smallskip}
            \rm{Date} & \rm{MJD} & \rm{Phase}^{a} & \rm{Exp. \;time} & q_{\rm{ave}} & u_{\rm{ave}} & \rm{Pol.\;degree} & \rm{Pol. \;angle} & \rm{Filter} \\
            (\rm{UT}) & (\rm{days}) & (\rm{days}) & (\rm{seconds}) & (\%) & (\%) & (\%) & (\rm{degree}) & \\
            \noalign{\smallskip}
            \hline\hline
            \noalign{\smallskip}
            \multirow{2}{*}{2021-10-05.13} & \multirow{2}{*}{59492.13} & \multirow{2}{*}{-6.33} & 4 \times 80 & -0.10 \pm 0.15 & -0.60 \pm 0.18 & 0.56 \pm 0.18 & 130.3 \pm 3.6 & V \\ 
             & & & 4 \times 80 & -0.12 \pm 0.05 & -0.55 \pm 0.14 & 0.53 \pm 0.14 & 128.8 \pm 1.5 & R \\
            \noalign{\smallskip} \hline \noalign{\smallskip}
            \multirow{2}{*}{2021-10-12.20} & \multirow{2}{*}{59499.20} & \multirow{2}{*}{+0.74} & 4 \times 50 & -0.20 \pm 0.13 & -0.35 \pm 0.10 & 0.37 \pm 0.11 & 120.1 \pm 4.4 & V \\
            & & & 4 \times 50 & -0.18 \pm 0.07 & -0.15 \pm 0.13 & 0.19 \pm 0.10 & 109.9 \pm 6.7 & R \\
            \noalign{\smallskip} \hline \noalign{\smallskip}
            \multirow{2}{*}{2021-11-09.93} & \multirow{2}{*}{59527.93} & \multirow{2}{*}{+29.47} & 4 \times 120 & 0.31 \pm 0.14 & 0.48 \pm 0.31 & 0.44 \pm 0.27 & 28.6 \pm 5.1 & V \\
            & & & 4 \times 80 & -1.08 \pm 0.57 & -1.60 \pm 0.24 & 1.86 \pm 0.38 & 118.0 \pm 3.6 & R \\
            \noalign{\smallskip} \hline \noalign{\smallskip}
            \multirow{2}{*}{2021-11-30.91} & \multirow{2}{*}{59548.91} & \multirow{2}{*}{+50.45} & 4 \times 120 & -0.29 \pm 0.54 & 0.45 \pm 0.37 & 0.19 \pm 0.43 & 61.4 \pm 13.3 & V \\
            & & & 4 \times 100 & -1.22 \pm 0.57 & -0.37 \pm 0.52 & 1.02 \pm 0.57 & 98.4 \pm 5.9 & R \\
            \noalign{\smallskip}
            \hline
         \end{array}
         $\\
         \begin{minipage}{.88\hsize}
        \smallskip
        Notes. ${}^{a}$Relative to the $B$-band peak: $t_{0}=59498.46$ \citep[MJD][]{Dimitriadis2023}.
        \end{minipage}
   \end{table*}


\begin{table*}
      \caption[]{Log of the imaging-polarimetric observations of SN 2022ilv and their measurements.}
      \label{tab:log_22ilv}
      $
         \begin{array}{lcccccccccc}
            \hline
            \noalign{\smallskip}
            \rm{Date} & \rm{MJD} & \rm{Phase}^{a} & \rm{Exp. \;time} & q_{\rm{ave}} & u_{\rm{ave}} & \rm{Pol.\;degree} & \rm{Pol. \;angle} & \rm{Filter} \\
            (\rm{UT}) & (\rm{days}) & (\rm{days}) & (\rm{seconds}) & (\%) & (\%) & (\%) & (\rm{degree}) & \\
            \noalign{\smallskip}
            \hline\hline
            \noalign{\smallskip}
            \multirow{2}{*}{2022-05-03.18} & \multirow{2}{*}{59702.18} & \multirow{2}{*}{-5.28} & 4 \times 80 & -1.00 \pm 0.03 & 1.50 \pm 0.07 & 1.80 \pm 0.06 & 61.8 \pm 0.4 & V \\
            & & & 4 \times 80 &  -0.97\pm 0.11 & 1.88 \pm 0.15 & 2.11 \pm 0.14 & 58.6 \pm 0.8 & R \\
            \noalign{\smallskip} \hline \noalign{\smallskip}
            \multirow{2}{*}{2022-05-14.00} & \multirow{2}{*}{59713.00} & \multirow{2}{*}{+5.54} & 4 \times 80 & -1.57 \pm 0.19 & 1.95 \pm 0.07 & 2.50 \pm 0.13 & 64.4 \pm 0.9 & V \\
            & & & 4 \times 80 & 0.14 \pm 0.57 & 1.16 \pm 0.40 & 1.03 \pm 0.40 & 41.6 \pm 7.0 & R \\
            \noalign{\smallskip} \hline \noalign{\smallskip}
            \multirow{2}{*}{2022-06-03.95} & \multirow{2}{*}{59733.95} & \multirow{2}{*}{+26.49} & 4 \times 100 & -1.27 \pm 0.17 & 1.26 \pm 0.30 & 1.76 \pm 0.24 & 67.6 \pm 2.0 & V \\
            & & & 4 \times 100 & -1.14 \pm 0.08 & 0.64 \pm 0.11 & 1.30 \pm 0.09 & 75.3 \pm 1.1 & R \\
            \noalign{\smallskip} \hline \noalign{\smallskip}
            \multirow{2}{*}{2022-06-26.96} & \multirow{2}{*}{59756.96} & \multirow{2}{*}{+49.50} & 4 \times 120 & -1.57 \pm 0.09 & 0.29 \pm 0.20 & 1.59 \pm 0.10 & 84.8 \pm 1.8 & V \\
            & & & 4 \times 100 & -1.23 \pm 0.19 & 1.83 \pm 0.58 & 2.09 \pm 0.49 & 62.0 \pm 2.3 & R \\
            \noalign{\smallskip} \hline \noalign{\smallskip}
            \multirow{2}{*}{2022-07-07.94} & \multirow{2}{*}{59767.94} & \multirow{2}{*}{+60.48} & 4 \times 120 & 0.55 \pm 0.45 & 0.49 \pm 0.16 & 0.57 \pm 0.35 & 20.8 \pm 6.3 & V \\
            & & & 4 \times 120 & 0.35 \pm 0.66 & 1.49 \pm 0.99 & 0.91 \pm 0.98 & 38.4 \pm 6.4 & R \\
            \noalign{\smallskip}
            \hline
         \end{array}
         $\\
         \begin{minipage}{.88\hsize}
        \smallskip
        Notes. ${}^{a}$Relative to the $g$-band peak: $t_{0}=59707.46$ \citep[MJD][]{Srivastav2023}.
        \end{minipage}
   \end{table*}


\section{Results and discussion}

   \begin{figure*}
   \centering
   \includegraphics[width=0.49\hsize]{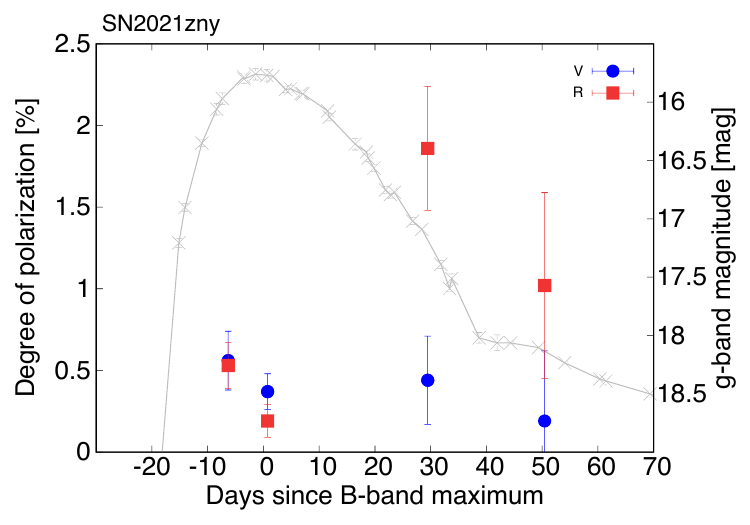}
   \includegraphics[width=0.49\hsize]{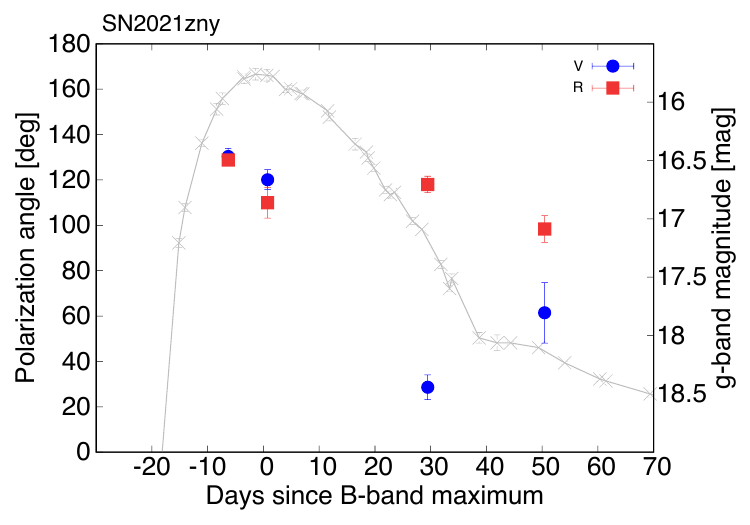}
      \caption{
      Polarization degree and angle of the $V$- (blue circle) and $R$-band (red square) polarization in SN~2021zny. The gray crosses connected with lines show the $g$-band LC of SN~2021zny from \citet[][]{Dimitriadis2023}.
              }
         \label{fig:fig1}
   \end{figure*}

    \begin{figure}
   \centering
   \includegraphics[width=\hsize]{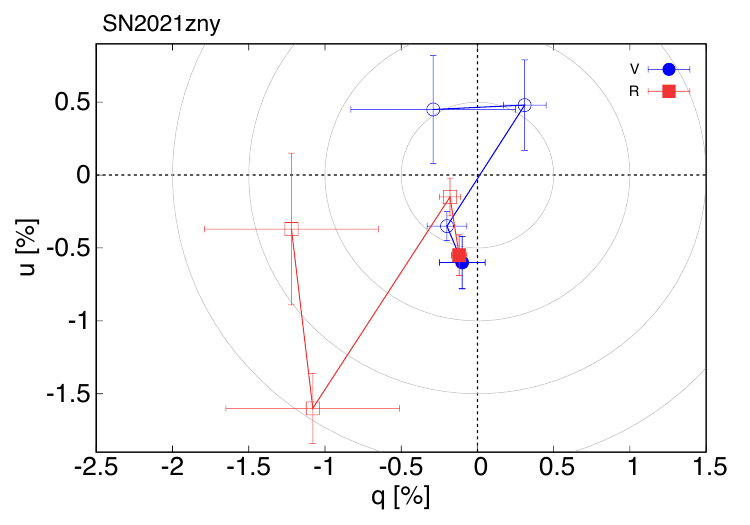}
      \caption{
      The time evolution of the $V$- (blue) and $R$-band (red) polarization in SN~2021zny in the $q-u$ plane. The first-epoch (Phase -6.33 days) data are shown with filled points. The gray lines correspond to the polarization degrees of 0.5, 1.0 and 1.5 \% from the origin, respectively.
              }
         \label{fig:fig2}
   \end{figure}

Figure~\ref{fig:fig1} shows the time evolution of the $V$- and $R$-band polarization of SN~2021zny. Both $V$- and $R$-band polarization shows $P\sim 0.4$ \% and $\theta \sim 120$ degrees around the peak. At Phase~+29.47 days, the degree of the $R$-band polarization is increased to $\sim 1.9$ \% keeping the polarization angle around $\sim 120$ degrees, followed by slightly smaller values of the polarization degree and angle at Phase~+50.45 days. On the other hand, the $V$-band polarization, at Phases~+29.47 and +50.45 days, shows small degrees of the polarization ($\lesssim 0.5$ \%) with different polarization angles ($\sim 45$ degrees) from $\sim 120$ degrees around the peak. The polarization is also shown in the $q-u$ plane (Figure~\ref{fig:fig2}).

The SN can have not only intrinsic polarization but also ISP. \citet[][]{Dimitriadis2023} estimated the total dust extinction in the Milky Way and in the host galaxy as $E(B-V)_{\rm{tot}}=0.14\pm0.07$ mag for SN~2021zny. The empirical relation by \citet[][]{Serkowski1975} indicates that its ISP should likely have $P_{\rm{max}} \lesssim 1.3$ \%. 
From the values of the polarization degree, the polarization component whose angle is $\sim 45$ degrees (the $V$-band polarization at Phases~+29.47 and +50.45 days) can be explained to be due to the ISP. This interpretation may naturally explain the discrepancy between the $V$- and $R$-band polarization at Phases~+29.47 and +50.45 days. The $V$-band polarization show the ISP due to the line depolarization of the strong continuum polarization, while the $R$-band polarization reflect the strong continuum polarization. Alternatively, the $V$-band polarization show the ISP due to the low intrinsic SN continuum polarization, while the $R$-band polarization shows a strong line polarization. In this case, the axis of the aspherical distribution of the elements for the strong line polarization should be the same with that of the overall ejecta geometry predicted by the continuum polarization at the first two phases.
It is noted that the directions of the magnetic fields, which are supposed to be aligned with the ISP angle \citep[e.g.,][]{Davis1951}, in a spiral galaxy tends to follow the directions of the spiral arms \citep[e.g.,][]{Beck2015}. Therefore, the polarization angle of this component ($\theta \sim 45$ degrees), which nicely corresponds to the structure of the host galaxy \citep[see Figure~1 in][]{Dimitriadis2023}, might support this interpretation. 
Adopting this component ($P\sim 0.3$ \% and $\chi \sim 45$ degrees) as the ISP, the $V$- and $R$-band polarization show both $\sim 1.0$ \% of the intrinsic polarization around the brightness peak and then $\sim 0$ and $\sim 2.0$ \%, respectively, at later phase. Even if this component is another component of the intrinsic SN polarization and the ISP is negligible, the intrinsic polarization is high (($\sim 1- \sim 2$ \%)). The high polarization is on the highest end of the diversity in Type~Ia SN polarization or possibly beyond \citep[$\lesssim 1$ \%, e.g.,][]{Cikota2019}. This indicates that the ejecta of SN~2021zny is significantly aspherical, even compared to the extreme cases of normal Type~Ia SNe.

It is noted that there is another possibility for the origin of the continuum polarization in Type~Ia SNe, i.e., the polarization due to the scattering by circumstellar dust \citep[e.g.,][]{Nagao2018,Hu2022}. The polarization created by this mechanism has wavelength dependence (typically higher polarization in bluer wavelengths) and time evolution (typically higher polarization at the beginning of the tail phase than at the peak) as demonstrated in \citet[][]{Nagao2018}. There are some observational examples \citep[e.g.,][]{Yang2018}. Firstly, the polarization in SN~2021zny does not show clear wavelength dependence at Phases~+29.47 and +50.45 days, while it shows higher degrees in the $R$ band than those in the $V$ band at latter phase. Secondly, SN~2021zny shows relatively high polarization degrees already before the B-band peak ($\sim 0.4$ \% at Phase~-6.33). These features cannot be explained with the dust scattering scenario. Therefore, we reject the possibility of scattering in circumstellar dust as the origin of the polarization in SN~2021zny.

   \begin{figure*}
   \centering
   \includegraphics[width=0.49\hsize]{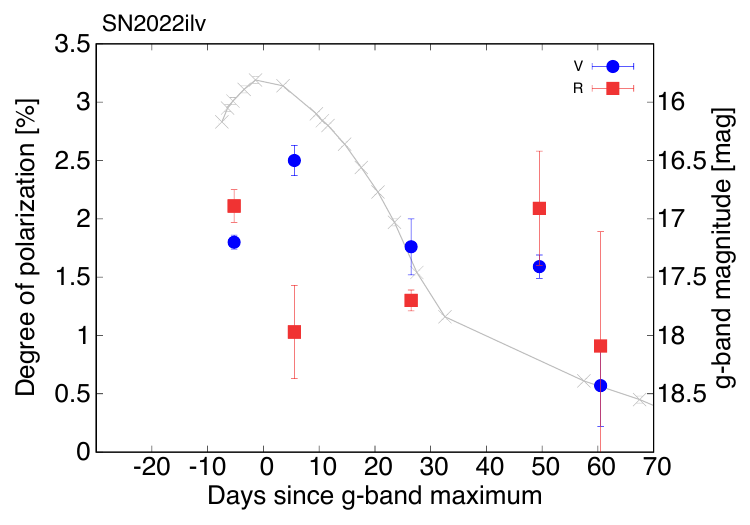}
   \includegraphics[width=0.49\hsize]{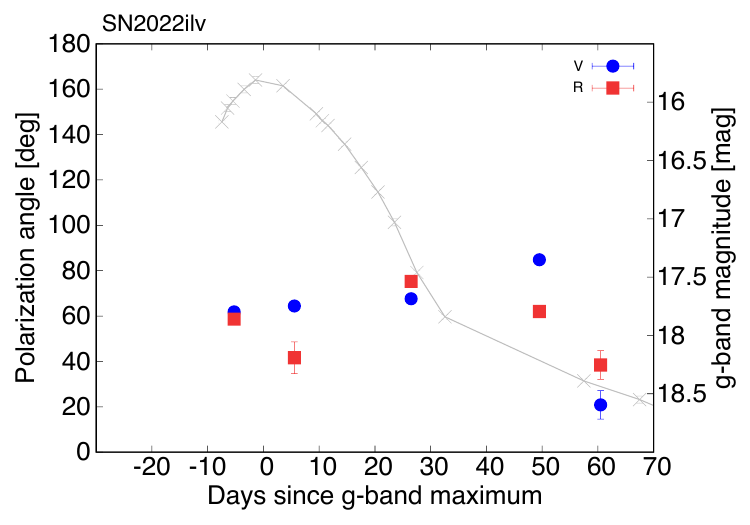}
      \caption{
      Same as Figure\ref{fig:fig1} but for SN~2022ilv. The gray dots connected with lines show the $g$-band LC of SN~2022ilv from \citet[][]{Srivastav2023}.
              }
         \label{fig:fig3}
   \end{figure*}

    \begin{figure}
   \centering
   \includegraphics[width=\hsize]{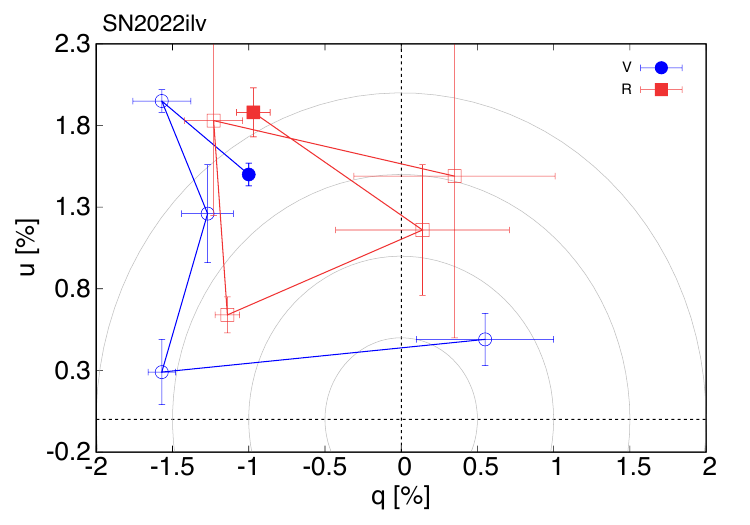}
      \caption{
      Same as Figure\ref{fig:fig2} but for SN~2022ilv. The gray lines correspond to the polarization degrees of 0.5, 1.0, 1.5 and 2.0 \% from the origin, respectively.
              }
         \label{fig:fig4}
   \end{figure}

Figure~\ref{fig:fig3} shows the time evolution of the $V$- and $R$-band polarization in SN~2022ilv. The polarization degrees are high and time-variable around $\sim 2.0$ \% with a relatively constant polarization angle of around $\sim 60$ degrees, except for the data at Phase +60.48 days ($P\sim 0.5$ \%, $\theta \sim 30$ degrees). The polarization degrees and angles in the $V$ and $R$ bands are relatively consistent at all phases except Phase~+5.54, indicating wavelength-independent polarization, i.e., continuum polarization rather than line polarization. The polarizaiton at Phase~+5.54 may be due to the effects of line polarization and depolarization. This behavior of the polarization is also seen in the $q-u$ plane (Figure~\ref{fig:fig4}), although the early-phase points are clustered around a point deviating from the points at Phase +60.48 days, except the $R$-band point at Phase~+5.54.

\citet[][]{Srivastav2023} estimated the dust extinction for SN~2022ilv to be $E(B-V)_{\rm{tot}}=0.11$ mag, assuming that the extinction arises only from the MW dust because the host galaxy is extremely faint and thus should have a low metalicity and small amount of dust. Adopting this value, the empirical relation by \citet[][]{Serkowski1975} indicates that its ISP should have $P_{\rm{max}} \lesssim 1.0$ \%. Given that the observed polarization around the peak is too high to be dominated by the ISP, it should be dominated by the intrinsic SN polarization. The polarization degrees at Phase +60.48 days can be consistent with the ISP. However, the polarization angle of the ISP in the MW along the line of sight to SN~2022ilv is estimated to be $\sim 130$ degrees by polarimetric observations of MW stars at $\sim 100- \sim600$ pc \citep[Figure~3 in][]{Berdyugin2014}. Given that the MW extinction for SN~2022ilv is mainly caused by dust at $\lesssim 140$ pc \citep[][]{Green2019}, the ISP can be estimated to be $\lesssim 1.0$ and $\sim 0.5$\% as an averaged value \citep[Figure~4 in][]{Berdyugin2014}. Therefore, we conclude that the ISP is negligible ($\lesssim 0.5$\%) and the polarization at Phase +60.48 days might still express another intrinsic component with a slightly different angle ($P\sim 0.5$ \%, $\theta \sim 30$ degrees), in addition to the intrinsic component at the early phases ($\sim 2.0$ \% and $\sim 60$ degrees). This might indicate inhomogeneous structures in the SN ejecta.
Even if we assume that the polarization at Phase +60.48 days is dominated by the ISP, the ISP-subtracted intrinsic SN polarization at early phases is high ($P\sim 2.0$ \%). It is noted that, as in the case of SN~2021zny, the polarization in SN~2021ilv also cannot be explained by the scattering in circumstellar dust.
In any case, the intrinsic SN polarization is very high $\sim 2$ \%, which is the highest intrinsic continuum polarization observed in any Type~Ia SN \citep[e.g.,][]{Wang2008,Cikota2019}. This implies that the ejecta of SN~2022ilv is also very aspherical, compared to any other Type~Ia SNe.

The $V$- and $R$-band polarization in SNe~2021zny and 2022ilv shows high degrees ($\sim1- \sim2$\%), indicating large aspherical structures. According to the numerical calculations of the polarization signal in various scenarios for Type~Ia SNe by \citet[][]{Bulla2016a,Bulla2016b},
the only possible scenario to show such high polarization, i.e., a large aspherical structure, is the violent merger scenario. Even though the aspherical structures in the 03fg-like Type~Ia SNe have been suggested using several different modes of observations (see Section~1), our observations provide the first evidence from polarimetry for such aspherical structures.

\begin{acknowledgements}
We thank Masayuki Yamanaka for helpful discussions.
This work is based on observations made under program IDs P64-023, P65-004 and P65-005 with the Nordic Optical Telescope, owned in collaboration by the University of Turku and Aarhus University, and operated jointly by Aarhus University, the University of Turku and the University of Oslo, representing Denmark, Finland and Norway, the University of Iceland and Stockholm University at the Observatorio del Roque de los Muchachos, La Palma, Spain, of the Instituto de Astrofisica de Canarias.
TN acknowledges support from the Research Council of Finland projects 324504, 328898 and 353019.
KM acknowledges support from the Japan Society for the Promotion of Science (JSPS) KAKENHI grant (JP20H00174 and JP24H01810) and by the JSPS Open Partnership Bilateral Joint Research Project between Japan and Finland (JPJSBP120229923).
S. M. was funded by the Research Council of Finland project 350458.
HK was funded by the Research Council of Finland projects 324504, 328898, and 353019.
CPG acknowledges financial support from the Secretary of Universities and Research (Government of Catalonia) and by the Horizon 2020 Research and Innovation Programme of the European Union under the Marie Sk\l{}odowska-Curie and the Beatriu de Pin\'os 2021 BP 00168 programme, from the Spanish Ministerio de Ciencia e Innovaci\'on (MCIN) and the Agencia Estatal de Investigaci\'on (AEI) 10.13039/501100011033 under the PID2020-115253GA-I00 HOSTFLOWS project, and the program Unidad de Excelencia Mar\'ia de Maeztu CEX2020-001058-M. 
\end{acknowledgements}

%
%

\bibliographystyle{aa} 
\bibliography{aa.bib}

\end{document}